\documentclass{article}

\usepackage{amssymb}
\usepackage{amsfonts}
\usepackage{latexsym}
\usepackage[T1,T2A]{fontenc}
\usepackage[cp1251]{inputenc}
\usepackage[T2B]{fontenc}
\usepackage[bulgarian,russian,english]{babel}

\newtheorem {theorem} {\bf Theorem}

\newtheorem {problem} {\bf Problem}
\newtheorem{task}{\bf Task}

\newtheorem{algorithm}{\bf Algorithm }
\newtheorem{proposition}{\bf Proposition}

\title {Some Combinatorial Problems on Binary Matrices in Programming Courses }
\author {Krasimir Yankov Yordzhev}
\date {}

\begin {document}
\inputencoding{cp1251}

\maketitle

\centerline{South-West University  ''N. Rilsky'', Blagoevgrad, Bulgaria}

\centerline{e-mail:  yordzhev@swu.bg; iordjev@yahoo.com}

\begin{abstract}
The study proves the existence of an algorithm to receive all elements of a class of binary matrices without obtaining redundant elements, e. g. without obtaining binary matrices that do not belong to the class. This makes it possible to avoid checking whether each of the objects received possesses the necessary properties. This significantly improves the efficiency of the algorithm in terms of the criterion of time.  Certain useful educational effects related to the analysis of such problems in programming classes are also pointed out.

\end{abstract}



\section{Introduction}\label{intr}

The stimulation of students' interest and motivation  a certain discipline or branch of science is a matter of combining a variety of methods. The present study discusses two techniques which experience has shown to be quite efficient in programming classes:
\begin{enumerate}
\item  Emphasizing the fact that the solution to a mathematical problem for all values of the parameters is yet to be discovered (for example problems \ref{z3} and \ref{probl2} below), students can be given an assignment to write a program solving the problem in certain specific cases, e. g. not for all the possible values of the parameters. In section \ref{prblms} two problems such as the ones described above will be formulated so that the students can write a program for relatively small values of the parameters $n$ and $k$.
\item It is also possible to point out that a particular problem may be solved by applying an algorithm that is more efficient as compared with standard algorithms which excelling students do not normally find difficult to apply. Section \ref{sect2} offers a specific example how to apply this technique.
\end{enumerate}

The present study is thus especially useful for students educated to become programmers as well as for their instructors and lecturers.

The paper discusses certain combinatorial problems on binary matrices.

For the classification  of all non defined concepts and notations
as well as for common assertion which have not been proved here,
we recommend sources \cite{aigner,lankaster,stan,baranov,tar2}.

\section{Some mathematical problems whose solution for all values of the parameters has not been discovered and certain results related to these problems}\label{prblms}

A \emph{binary} (or   \emph{boolean}, or (0,1)-\emph{matrix})
 is a matrix whose all elements belong to the set ${\bf B}
=\{ 0,1 \}$. With ${\bf B}_n$ we will denote the set of all  $n
\times n$  binary matrices.

Using the notation from   \cite {1}, we will call
$\Lambda_n^k$-matrices all $n\times n$ binary matrices in each row
and each column of which there are exactly $k$ in number 1's.

\begin{problem}\label{z3} Find out the number of all   $n\times n$ binary
matrices containing exactly  $k$ elements equal to 1 in each row and each column,
e.g. the number of all $\Lambda_n^k$-matrices.
\end{problem}

Let us denote the number of all $\Lambda_n^k$-matrices with
$\lambda (n,k)$.

Problem  \ref{z3} has not been solved for all values of the parameters. That is there is no known formula to calculate  the $\lambda (n,k)$  for all $n$ and $k$. There are formulas for the calculation of the function $\lambda (n, k)$ for each $n$ for relatively small values of $k$; more specifically, for $k = 1$, $k = 2$ and $k = 3$. We do not know formula to calculate the function $\lambda (n, k)$  for $k > 3$ and for all positive integer $n$.

It is easy to prove the following well-known formula:

\begin{equation}\label{permmmm}
\lambda (n,1) =n!
\end{equation}

In \cite{6} is offered the formula:

\begin{equation}\label{1}
\lambda (n,2) = \sum_{2x_2 +3x_3 + \cdots +nx_n =n}
\frac{(n!)^2}{\displaystyle \prod_{r=2}^n x_r !(2r)^{x_r}}
\end{equation}

One of the first recursive formulas for the calculation of
$\lambda (n,2)$ appeared in \cite{anand} (see also \cite[p. 763]{gupta}):
\begin{equation}\label{2}
\begin{array}{l}
\displaystyle \lambda (n,2) =\frac{1}{2} n(n-1)^2 \left[ (2n-3)
\lambda (n-2,2) +(n-2)^2 \lambda (n-3,2) \right]\; ; \quad   n\ge 4 \\
\displaystyle \lambda (1,2) =0,\quad \lambda (2,2) =1,\quad
\lambda (3,2) =6
\end{array}
\end{equation}

Another recursive formula for the calculation of $\lambda (n,2)$
occurs  in \cite{13} :
\begin{equation}\label{3}
\begin{array}{l}
\displaystyle \lambda (n,2) = (n-1)n\lambda (n-1,2)
+\frac{(n-1)^2
n}{2} \lambda (n-2,2)\; ;  \quad n\ge 3 \\
\lambda (1,2) =0,\quad \lambda (2,2) =1
\end{array}
\end{equation}

The following recursive system  for the calculation of
$\lambda (n,2)$ is put forward in \cite{iord}:
\begin{equation}\label{4}
\left|
\begin{array}{l}
\lambda (n+1,2) =n(2n-1)\lambda (n,2) + n^2 \lambda (n-1,2) - \pi(n+1) \; ; \quad n\ge 2\\
 \pi (n+1) = \\
 = {n^2 (n-1)^2 \over 4} [8(n-2)(n-3) \lambda (n-2,2) +(n-2)^2 \lambda (n-3,2) -4\pi (n-1)]\; ; \quad n\ge 4 \\
\lambda (1,2) = 0 ,\quad \lambda (2,2) = 1 ,\quad \pi (1) = \pi (2) =
\pi (3) = 0 ,\quad \pi (4) = 9
\end{array}
\right.
\end{equation}
where  $\pi (n)$ identifies the number of a special class of
$\Lambda_n^2$-matrices.

The following formula in an explicit form for the calculation of
$\lambda (n,3)$ is offered in \cite{stein}.

\begin{equation}\label{111}
 \lambda (n,3)=\frac{n!^2}{6^n} \sum \frac{(-1)^\beta (\beta
+3\gamma )! 2^\alpha 3^\beta}{\alpha !\beta ! \gamma !^2 6^\gamma}
\end{equation}
where the sum is done as regard all $ \frac{(n+2)(n+1)}{2}$
solutions in nonnegative integers of the equation   $\alpha
+\beta+\gamma =n$.
As it is noted in  \cite{stan} formula    (\ref{111}) does not
give us good opportunities to study behavior of  $\lambda (n,3)$.

Let $n$ be a positive integer and let $A\in {\bf B}_{n^2}$ is a $n^2 \times n^2$ binary matrix.  With the help of $n - 1$ horizontal lines and $n - 1$ vertical lines $A$ has been divided into $n^2$ of number non-intersecting $n\times n$ square sub-matrices $A_{kl}$, $1\le k,l\le n$, e. g.

\begin{equation}\label{matrA}
A =
\left[
\begin{array}{cccc}
A_{11} & A_{12} & \cdots & A_{1n} \\
A_{21} & A_{22} & \cdots & A_{2n} \\
\vdots & \vdots & \ddots & \vdots \\
A_{n1} & A_{n2} & \cdots & A_{nn}
\end{array}
\right] .
\end{equation}
The sub-matrices $A_{kl}$, $1\le k,l\le n$ will be called blocks.

Adding one more condition, we can make the problem \ref{z3} more complicated:

\begin{problem} \label{probl2}

Find out the number $\mu (n,k)$ of all $n^2 \times n^2$  binary matrices that have $k$ elements equal to 1 in each row, each column, and each $n\times n$ block.

\end{problem}

As demonstrated in  \cite{dahl}, problem \ref{probl2} has to do with the solution of a variety of combinatorial problems associated with the Sudoku riddles.

Problem \ref{probl2} is solved in  \cite{dahl} for $k = 1$ and in \cite{yordzhev} other methods are used to prove that

\begin{equation}\label{sssssss}
\mu (n,1)  =\left( n! \right)^{2n}
\end{equation}

No formula has been put forward for the calculation of the function $\mu (n,k)$ when $k > 1$.

\section{S-permutation matrices}\label{sect2}

 A matrix $A\in {\bf B}_{n^2}$ is called S-\emph{permutation} if in each row, each column, and each block, of $A$ there is exactly one 1. Let the set of all $n^2 \times n^2$ S-permutation matrices be denoted by $\Sigma_{n^2}$.

Two   matrices $A=(a_{ij} )\in \Sigma_{n^2}$ and $B=( b_{ij} )\in \Sigma_{n^2}$, $1\le i,j\le n^2$ will be called \emph{disjoint}, if there are not elements with one and the same indices $a_{ij}$ and $b_{ij}$ such that $a_{ij} =b_{ij} =1$.

The following obvious proposition is given in \cite{dahl}:

\begin{proposition}\label{disj} \cite{dahl}
Square $n^2 \times n^2$ matrix $P$ with elements of $\mathbb{Z}_{n^2} =\{ 1,2,\ldots ,n^2 \}$ is Sudoku matrix if and only if there are  matrices $A_1 ,A_2 ,\ldots ,A_{n^2} \in\Sigma_{n^2}$, each two of them are disjoint and such that $P$ can be given in the following way:
$$P=1\cdot A_1 +2\cdot A_2 +\cdots +n^2 \cdot A_{n^2}$$
\hfill $\Box$
\end{proposition}

Let us analyze the following programming task:

\begin{task}\label{task1}
 Write a program to obtain all S-permutation $n^2 \times n^2$ matrices for a specific positive  integer $n$.
\end{task}

Experience shows that the majority of students do not fined it difficult to solve a task such as the one offered above. Unfortunately, the solutions they normally suggest are not very efficient. Below we present of the most common solutions given by students:

It is easy to observe that if we remove the condition to have only one 1 for each block of the $n^2 \times n^2$ binary matrices, the task above can be transformed into a task for the obtaining of all permutations of the integers from 1 to $n^2$. This combinatorial task is often discussed in programming classes and a clear-cut solution can be found in a number of study books, such as  \cite{nakov}. Let $\pi =\langle p_1 ,p_2 ,\ldots , p_m \rangle $ be a permutation of the integers from 1 to $m$. Then we obtain the  $m\times m$ binary matrix $B=(b_{ij} )\in {\bf B}_m$, such that $b_{ij} =1$ if and only if $p_i =j$, $1\le i,j\le m$. It is clear that the matrix $B$ obtained in this case has one 1 in each row and each column. This is where the name of such matrices comes from: \emph{permutation matrices}. This gives us the following algorithm for the solution to task \ref{task1}:

\begin{algorithm}\label{alg1}Obtaining all S-permutation matrices.
\begin{enumerate}
\item \label{i1} Obtain all the permutations of integers from 1 to $n^2$;
\item \label{i2} For each permutation $\pi =\langle p_1 ,p_2 ,\ldots , p_{n^2} \rangle $  obtained in step \ref{i1}, obtain the binary matrix $A=(\alpha_{ij} )\in {\bf B}_{n^2}$,  such that  $\alpha_{ij} =1$ if and only if  $p_i =j$. In all other cases $\alpha_{ij} =0$, $1\le i,j\le n^2$;
\item \label{i3} For each matrix obtained in step  \ref{i2}, check whether each block has only one 1. If ({\bf true}) then the matrix is S-permutation, if ({\bf false}) then we remove this matrix from the list.
\end{enumerate}
\end{algorithm}

Unfortunately, algorithm \ref{alg1} entails the obtaining of a variety of redundant matrices and a lot of time is wasted to check whether these meet the conditions (step \ref{i3}). The total number of the permutation matrices obtained in algorithm \ref{alg1} is $n^2!$, while according to formula (\ref{sssssss}) the number of the S-permutation matrices is $(n!)^{2n} $. But it is not difficult to see that   $n^2 ! >(n!)^{2n}$ when $n \ge 2$. Importantly, the program implementation in step \ref{i3} of algorithm \ref{alg1} is also significantly aggravated and requires certain efforts and mathematical competence.

When $n = 2$, we have $2^2 !=24$,  $(2!)^{2\cdot 2} =16$; when $n = 3$, we have $3^2 !=362\; 880$,  $(3!)^{2\cdot 3} =46\; 656$; when $n = 4$, we have  $4^2 !=20\; 922\; 789\; 888\; 000$, $(4!)^{2\cdot 4} =110\; 075\; 314\; 176$, etc. It is possible to prove that as $n$ increases, the value of expression  $\displaystyle \frac{n^2 !}{(n!)^{2n}}$ increases as well. This proves the inadequate efficiency of algorithm  \ref{alg1}.

The present study will demonstrate that there is an algorithm for the obtaining of all S-permutation matrices which bypasses the redundant, non S-permutation matrices. This reduces the iterations in the algorithm to the absolute minimum and each iteration bypasses the checking whether the matrix obtained is S-permutation. Such an algorithm is obviously more efficient and takes less time to apply than algorithm \ref{alg1}. It is based on theorem \ref{ThPin} proven below:

Let denote with $\Pi_n$ the set of all $(2n)\times n$ matrices, which we shortly call $\Pi_n$ matrices, in which every row is a permutation of all elements of $\mathbb{Z}_n =\{ 1,2,\ldots ,n\}$. It is obvious that
\begin{equation}\label{Pi_n}
\left| \Pi_n \right| =\left( n! \right)^{2n}
\end{equation}

We will give a little bit more complicated definition of the term disjoint about $\Pi_n$ matrices. Let $C=(c_{ij} )$ and $D=(d_{ij} )$ be two $ \Pi_n$ matrices. We say that $C$ and $D$ are \emph{disjoint}, if there are not natural numbers $s,t\in \{ 1,2,\ldots n\}$ such that ordered pair $\langle c_{st} ,c_{n+t\; s} \rangle$ has to be equal to the ordered pair $\langle d_{st} ,d_{n+t\; s} \rangle$.

\begin{theorem}\label{ThPin}
There is a bijective map from $\Pi_n$ to $\Sigma_{n^2}$ and the pair of disjoint matrices of $\Pi_n$ corresponds to the pair of disjoint matrices of $\Sigma_{n^2}$.
\end{theorem}

Proof. Let $P=(p_{ij} )_{2n\times n} \in \Pi_n$. We obtain the unique matrix of $\Sigma_{n^2}$ from $P$ with the help of the following algorithm:

\begin{algorithm}\label{alg5}
Obtaining  just one matrix of $\Sigma_{n^2}$ if there is given $P=(p_{ij} )_{2n\times n} \in \Pi_n$.
\begin{enumerate}
\item \textbf{for}  $ s = 1,2,\ldots ,n$ \textbf{do}
\item \textbf{for}  $ t = 1,2,\ldots ,n$ \textbf{do}

\textbf{begin}

\item $k:=p_{st}$;
\item $l:=p_{n+t\; s}$;
\item Obtain $n\times n$ matrix $A_{st} =(a_{ij})_{n\times n}$ such that $a_{kl}= 1$ и $a_{ij} =0$ in all other occasions;

\textbf{end};
\item Obtain matrix $A$ according to formula (\ref{matrA});

\end{enumerate}
\end{algorithm}

Let $s\in \mathbb{Z}_n =\{ 1,2,\ldots ,n\}$. Since ordered $n$-tuple $\langle p_{s1} ,p_{s2} ,\ldots ,p_{sn} \rangle$ which is $s$-th row of the matrix $P$ is a permutation, then in every row of $n\times n^2$ matrix
$$
R_s =
\left[
\begin{array}{cccc}
A_{s1} & A_{s2} & \cdots & A_{sn}
\end{array}
\right]
$$
there is only one 1. For every $j=1,2,\ldots ,n$ $A_{sj}$ is binary $n\times n$ matrix in this case. Analogously for every $t\in \mathbb{Z}_n$ because ordered $n$-tuple $\langle p_{n+t\; 1} ,p_{n+t\; 2} ,\ldots ,p_{n+t\; n} \rangle$ which is $(n+t)$-th row of $P$ is a permutation, then in every column of $n^2 \times n$ matrix
$$
C_t =
\left[
\begin{array}{c}
A_{1t}  \\
A_{2t} \\
\vdots \\
A_{nt}
\end{array}
\right]
$$
there is only one 1, where $A_{it}$, $i=1,2,\ldots ,n$ is a binary $n\times n$ matrix. Hence, the matrix $A$ which is obtained with the help of algorithm \ref{alg5} is $\Sigma_{n^2}$ matrix.

Since for every $P\in \Pi_n$ with the help of algorithm \ref{alg5} is obtained unique element of $\Sigma_{n^2}$, then this algorithm is a description of the map $\varphi \; :\; \Pi_n \to \Sigma_{n^2}$. It is easy to see that if there are given different elements of $\Pi_n$, with the help of algorithm \ref{alg5} we can obtain different elements of $\Sigma_{n^2}$. Hence, $\varphi$ is an injection. But according to formulas (\ref{sssssss}) and   (\ref{Pi_n}) $\left| \Sigma_{n^2} \right| =\left| \Pi_n \right|$, from where it follows that $\varphi$ is a bijection.

Analyzing algorithm \ref{alg5} we take the conclusion that $P$ and $Q$ are disjoint matrices of $\Pi_n$ if and only if $\varphi (P)$ and $\varphi (Q)$ are disjoint matrices of $\Sigma_{n^2}$ according to the above given definitions. The theorem is proved.

\hfill $\Box$

As an entailment of theorem \ref{ThPin}, the following  algorithm is received for the obtaining of all S-permutation matrices, which, based on the arguments above, can be claimed to be considerably more efficient than algorithm  \ref{alg1} for the same problem.

\begin{algorithm}\label{efectalg} Getting  of all S-permutation matrices.
\begin{enumerate}
\item \label{m1}Obtain all the permutations of integers from 1 to $n^2$;
\item \label{m2} With the help of all permutations obtained in step  \ref{m1}, obtain all $\Pi_n$ matrices;
\item\label{m3} From each $\Pi_n$  matrix obtained in step  \ref{m2}, obtain the next S-permutation matrix with the help of algorithm  \ref{alg5}.
\end{enumerate}
\end{algorithm}

\vspace{0.3cm}
Associate Professor Dr. Krasimir Yankov Yordzhev

South-West University  ''N. Rilsky''

2700 Blagoevgrad, Ivan Mihaylov Str. 66

Bulgaria

\rm e-mail:  yordzhev@swu.bg; iordjev@yahoo.com

\vspace{0.3cm}
\begin{center}
{\bf Некоторые комбинаторные задачи с бинарными матрицами на курсах программирования}
\end{center}

Исследование доказывает существование алгоритма для получения всех эле\-мен\-тов клас\-са бинарных матриц без получения избыточных эле\-мен\-тов, т.е. без по\-лу\-че\-ния би\-нар\-ных мат\-риц, которые не принадлежат к этому классу. Это дает воз\-мож\-ность из\-бе\-жать проверки, обладает ли каждый из полученных объек\-тов тре\-бу\-е\-мы\-ми свой\-ства\-ми. Так во много раз улучшается эффективность алгоритма в связи с кри\-те\-ри\-ем времени.  Обращается внимание на выгоды из рассмотреных задач для обу\-че\-ния по прог\-рам\-ми\-ро\-ва\-нию.

{\it \textbf{Ключевые слова:} стимулирование интерес студентов, мотивация к обу\-че\-нию, подготовка в области программирования, бинарная матрица, S-матрица пе\-ре\-ста\-но\-вок, комбинаторный алгоритм}

\vspace{0.5cm}
Доц. д-р Красимир Я. Йорджев

Юго-западный университет ''Неофит Рилски''

2700 Благоевград, ул. ''Иван Михайлов'' № 66

Болгария

\rm e-mail:  yordzhev@swu.bg; iordjev@yahoo.com

\end{document}